\documentclass[floatfix,reprint,aip,apl,longbibliography,twocolumn]{revtex4-2} 
\usepackage{lipsum}
\usepackage{graphicx}
\usepackage{amsmath}
\usepackage{amssymb}
\usepackage{epsfig}
\usepackage{subeqnarray}
\usepackage{color}
\usepackage{bbold}
\usepackage{dsfont}
\usepackage{tikz}
\usepackage{cancel}
\usetikzlibrary{positioning,calc}
\usepackage[version=3]{mhchem}
\usepackage{color}
\usepackage[normalem]{ulem}

\newcommand{\mt}[1]{\mathrm{#1}}

\newcommand{\fig}[1]{Figure~\ref{#1}}

\newcommand{\degreeC}[1]{$\mathrm{#1^{\circ}C}$}

\newcommand{\ie}{\textit{i.e.},~}
\newcommand{\via}{\textit{via~}}
\newcommand{\eg}{\textit{e.g.,~}}

\newcommand{\mum}{~$\mt{\mu m}$~}

\newcommand{\cm}{$\mt{cm^{-1}}$~}
\newcommand{\pca}{$Pca2_1$}
\newcommand{\hfo}{\ce{HfO_2}}
\newcommand{\hzo}{\ce{Hf_{0.5}Zr_{0.5}O_2}~}
\graphicspath{{./Figures/}}

\begin{document}

\title{Photoinduced Patterning of Oxygen Vacancies to Promote the Ferroelectric Phase of \ce{Hf_{0.5}Zr_{0.5}O_2}}

\author{Thomas E. Beechem}
\email[Authors to whom correspondence should be addressed: ]{tbeechem@purdue.edu, jihlefeld@virginia.edu}
\author{Fernando Vega}
\affiliation{School of Mechanical Engineering and Birck Nanotechnology Center, Purdue University, West Lafayette, IN 47907, USA}
\author{Samantha T. Jaszewski}
\author{Benjamin L. Aronson}
\affiliation{Department of Materials Science and Engineering, University of Virginia, Charlottesville, VA 22904, USA}
\author{Kyle P. Kelley}
\affiliation{Center for Nanophase Materials Sciences, Oak Ridge National Laboratory, Oak Ridge, TN 37381, USA}
\author{Jon. F. Ihlefeld}
\email[Authors to whom correspondence should be addressed: ]{tbeechem@purdue.edu, jihlefeld@virginia.edu}
\affiliation{Department of Materials Science and Engineering, University of Virginia, Charlottesville, VA 22904, USA}
\affiliation{Charles L. Brown Department of Electrical and Computer Engineering, University of Virginia, Charlottesville, VA 22904, USA}
\date{\today}

\begin{abstract}
Photoinduced reductions in the oxygen vacancy concentration were leveraged to increase the ferroelectric phase fraction of \hzo (HZO) thin-films. Modest ($\sim 0.02-0.77~\mathrm{mJ/\mu m^2}$) laser doses of visible light (488 nm, 2.54 eV) spatially patterned the concentration of oxygen vacancies as monitored by photoluminescence imaging.  Local, tip-based, near-field, nanoFTIR measurements showed that the photoinduced oxygen vacancy concentration reduction promoted formation of the ferroelectric phase (space group \pca) resulting in an increase in the piezoelectric response measured by piezoresponse force microscopy. Photoinduced vacancy tailoring provides, therefore, a spatially prescriptive, post-synthesis, and low-entry method to modify phase in \hfo-based materials.
\end{abstract}
\maketitle

Ferroelectric hafnium oxide\cite{boscke_2011} and alloyed counterparts like hafnium-zirconium oxide (\ce{Hf_{0.5}Zr_{0.5}O_2}, HZO)\cite{muller_2012} are being pursued for applications ranging from non-volatile memory to neuromorphic computing,\cite{schroeder_2022a,covi_2022} as they are scalable\cite{cheema_2020} to the length-scales of next generation computing architectures and are CMOS compatible.\cite{ihlefeld_2022}  These applications are reliant on the material's polar orthorhombic phase (space group \pca), which is responsible for its ferroelectric response.  This phase is metastable, however, necessitating means of both promoting its creation and stabilization.  Several approaches have been employed to this end including the incorporation of dopants,\cite{muller_2011a,muller_2011,schroeder_2018}  strain,\cite{shiraishi_2016,zhou_2022a}, top-electrode capping,\cite{boscke_2011,fields_2022} and tailoring the concentration of oxygen vacancies.\cite{jaszewski_2022,mittmann_2020,pal_2017}

Oxygen vacancies strongly influence the functional properties of \hfo-based devices because, depending on concentration, they can either stabilize the ferroelectric phase or induce transformations away from it.\cite{jan_2023,nukala_2021a,jaszewski_2022}  A classic ``Goldilocks problem," sufficient concentration of oxygen vacancies promotes formation of the ferroelectric phase but either ``too much" or ``too little" pushes the system towards other non-equilibrium, non-ferroelectric phases such as the tetragonal (space group $P42/nmc$) phase or its equilibrium monoclinic (space group $P2_1/c$) form.\cite{jaszewski_2022,hoffmann_2015} Maintaining a proper balance of oxygen vacancies is therefore necessary to enhance the performance and lifetime of \hfo-based devices. 

Realizing the balance of oxygen vacancies is complicated, however, since they are often charged.\cite{chai_2022} As such, when subjected to the requisite electric fields of device operation, charged oxygen vacancies migrate.\cite{nukala_2021a} The field-induced movement alters the local vacancy concentration in ways that can induce both phase transformation and the creation of shallow trap states within the bandgap that can cause current leakage.\cite{mallick_2023,cheng_2022b,nukala_2021a,hamouda_2022,jan_2023} Both effects limit \hfo-based device endurance and memory properties such as remanent polarization and retention.\cite{mallick_2023,alcala_2023} Oxygen vacancies can also be added or subtracted from \hfo\textemdash as opposed to just shuffled around\textemdash depending on the materials adjacent to it and their ability to getter or supply oxygen. This is one reason that the performance of \hfo-devices depends upon the electrodes used to supply the fields.\cite{fields_2021a,cao_2018a,tasneem_2022,hamouda_2020} Simply stated, oxygen vacancies are a prime factor dictating the performance of \hfo-based devices.

The impact of oxygen vacancies on the characteristics of \hfo~ is not solely detrimental.  Vacancies provide an additional knob for materials design. Specifically, extrinsically tailoring the vacancy concentration provides a way to increase the ferroelectric phase fraction and improve device performance.  The surfaces of thin films present an appealing opportunity to this end. Vacancies tend to cluster at surfaces and can be created and passivated there with relative ease.  Photoinduced phenomena are known to alter vacancy concentrations and stoichiometry at oxide surfaces.\cite{fujishiro_2005,glass_2019,jeong_2013,bharadwaja_2016}  These photoinduced modifications in defect concentration, for example, can induce a transformation from the anatase to rutile phase in \ce{TiO_2}.\cite{stagi_2015}  Here, the ability to promote the formation of the ferroelectric phase in HZO via photoinduced changes in oxygen vacancy concentration is demonstrated, thereby presenting a spatially deterministic, post-synthesis, low-entry method to tailor phase in this emergent material class. 

Ferroelectric \hzo was synthesized following methods reported previously.\cite{fields_2021d} Briefly,  100 nm-thick planar \ce{TaN} bottom electrodes were deposited on (001)-oriented silicon substrates using DC magnetron sputtering from a \ce{TaN} target with a power density of 3.3 $\mathrm{W/cm^2}$ within an argon background pressure of 5 mTorr. Plasma-enhanced atomic layer deposition (PE-ALD) was used to synthesize the 20 nm thick HZO films with an Oxford FlexAL II system operating at a temperature of \degreeC{260} with tetrakis(ethylmethylamido)hafnium (TEMA Hf) and tetrakis(ethylmethylamido)zirconium (TEMA Zr) precursors and an oxygen plasma as the oxidant.  Supercycles consisting of 6 cycles of \hfo~ and 4 cycles of \ce{ZrO_2} were utilized for deposition, resulting in a nominal film composition of \ce{Hf_{0.5}Zr_{0.5}O_2}. Subsequently, a 20 nm thick \ce{TaN} capping layer was deposited atop the \hzo to provide an out-of-plane mechanical constraint necessary for crystallization into the metastable ferroelectric phase during the rapid thermal anneal that followed.\cite{fields_2022} Conditions identical to that used for the bottom electrode deposition were employed. Films were rapid thermal annealed (RTA) at \degreeC{600} using an Allwin21 AccuThermo 610 Rapid Thermal Processor for 30 seconds in a \ce{N_2} atmosphere, whereupon the capping layer was removed with an SC1 etch owing to its high selectivity between \hfo~ and \ce{TaN}.\cite{hussain_2005,lomenzo_2015}  This etch exposed the \hzo surface.  Standard photolithographic processes were then performed to pattern gold fiducial marks atop the \ce{Hf_{0.5}Zr_{0.5}O_2} using electron beam evaporation. The fiducial marks allowed for co-location of scanning probe characterization techniques described later.

\begin{figure}[htbp]
\centering
\includegraphics[width=88 mm]{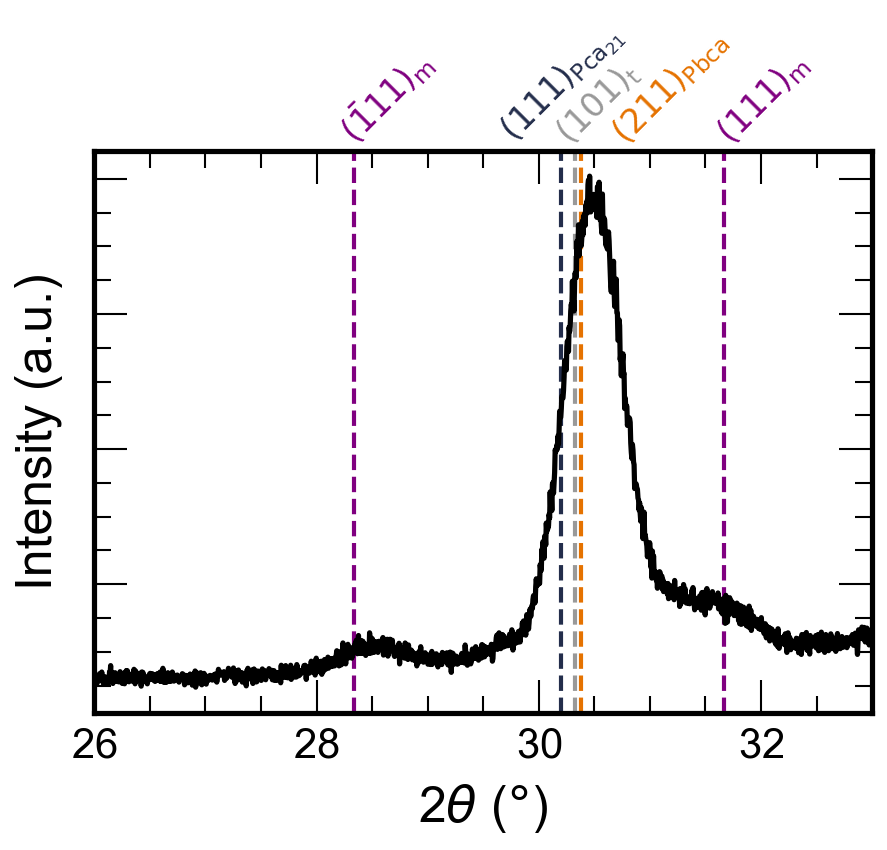}
\caption{Grazing-incidence X-ray diffraction pattern for the \ce{Hf_{0.5}Zr_{0.5}O_2} film.  The major feature at $30.5^{\circ}$ is indexed to a combination of the orthorhombic and tetragonal phases. The other features observed at $28.5^{\circ}$ and $31.4^{\circ}$ arise due to the monoclinic phase that makes up $14\%$ of the film. Vertical dashed lines are the indexed positions of \ce{HfO_2} and do not account for the effect of alloying or strain.}
\label{Fig_1}
\end{figure}

The phase make-up of the \hzo films was analyzed using grazing incidence X-ray diffraction (GIXRD) performed on a Panalytical Empyrean diffractometer in a pseudo-parallel beam configuration with a fixed incident angle of $0.7^\circ$.  The resulting diffraction pattern is shown in \fig{Fig_1}, where the dominant signature observed near $30.5^\circ$ can be indexed to a combination of the: polar orthorhombic (\ie ferroelectric, \pca), antipolar orthorhombic ($Pbca$), and tetragonal ($P4_2/nmc$) phases. Definitive differentiation between these three phases is difficult using XRD alone.\cite{jaszewski_2023,tharpe_2021} Signatures of the monoclinic phase ($P2_1/c$) are present but at comparatively much lower levels. The two peaks at $28.5^{\circ}$ and $31.4^{\circ}$, for example, are indexed as the ($\bar{\mathrm{1}}\mathrm{11}$) and (111) monoclinic reflections, respectively. By taking the ratio of integrated intensity from the main metastable feature at $30.5^{\circ}$ to the total integrated intensity of all three features, the volume fraction of monoclinic phase present in the sample was estimated to be $14\%$, similar to other HZO films reported by our team.\cite{fields_2022}  This phase fraction analysis assumes insignificant crystallographic texture and similar multiplicities and structure factors for the possible phases, which is a reasonable assumption based upon prior work.\cite{fields_2021d}  Subsequent nanoFTIR and piezoresponse force microscopy (PFM) measurements corroborate this finding by showing that the film is composed primarily of the polar and antipolar orthorhombic phases. 

\begin{figure*}[htbp]
\centering
\includegraphics[width=160 mm]{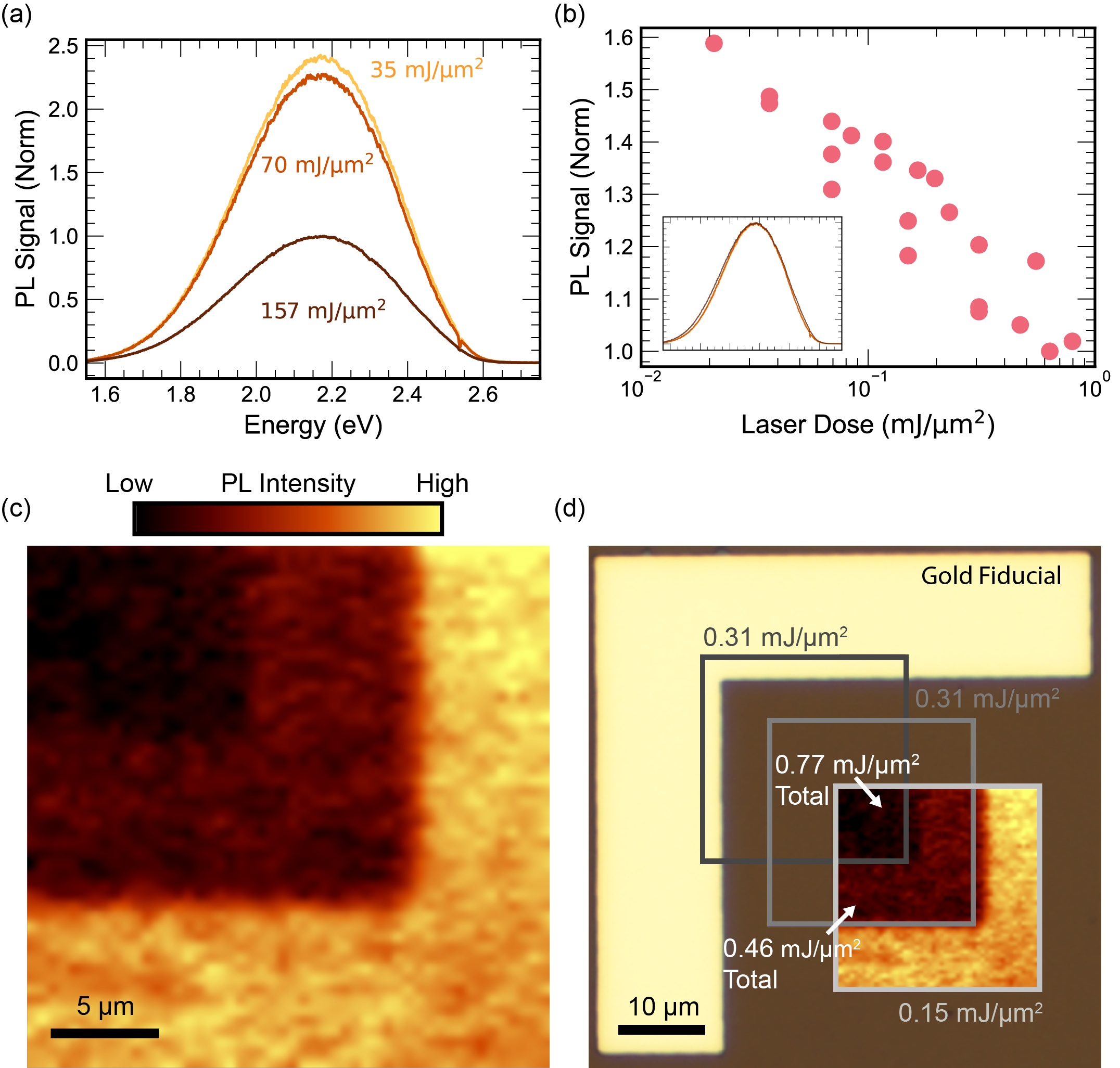}
\caption{(a) Normalized photoluminescence spectra of \ce{Hf_{0.5}Zr_{0.5}O_2} film acquired with differing laser powers where the signal near 2 eV is indicative of oxygen vacancies. (b) The PL signal decreases exponentially with laser dose and can be spatially patterned. (inset) Spectral data of (a) normalized so that baseline to peak is equal for all spectra. Overlapping profiles indicate that the oxygen-vacancy defect-state energies are not changing appreciably due to the laser.  (c) PL-image exhibiting reductions of intensity where laser dose is highest.  (d) Dose was changed by overlapping regions of successive PL images as delineated in the annotated optical image with the overlaid PL-image of (c). The gold fiducial was used to co-locate PL, nanoFTIR, and PFM measurements.}
\label{Fig_2}
\end{figure*}

Photoluminescence (PL) imaging was performed within the ambient atmosphere to assess the presence of oxygen vacancies and their modification with laser exposure. A Witec alpha300R system with a 488 nm laser focused to a diffraction limited spot size of $\sim 540$ nm with a 50X/0.55 NA objective was used. As shown in \fig{Fig_2}, a series of partially overlapping 20 by 20 \mum regions were successively imaged while changing the laser power.  For all PL-images, the sample was rastered beneath the laser at a speed of $5~\mathrm{\mu m/s}$ with a spectrum acquired every 0.1 s (\ie acquisition every 0.5 $\mu m$). The areal laser dose ($E$) was calculated \via $E = (4P\cdot t)/(\pi d^2)$, where $P$ is the power of the incident laser and $t$ the time required to move the sample a distance equal to the diffraction limited beam diameter, $d$. To compare the strength of the PL-response between differing laser powers, all signals were baseline subtracted and normalized relative to the number of photons that irradiated the sample. 
The PL spectrum of \hzo is shown in \fig{Fig_2}(a). A strong, slightly asymmetric, response centered near 2.13 eV is observed.  Photoluminescence in this region of the spectrum arises due to transitions between deep- and shallow-level trap states within the \hzo that emerge with the presence of oxygen vacancies.\cite{chai_2022,mallick_2023} Monitoring the strength of the photoluminescence signature, therefore, provides a means of quantifying the relative concentration of oxygen vacancies within the film. A stronger PL-signal indicates a higher oxygen-vacancy concentration for the ranges studied here.

As the laser dose increases, the relative strength of the PL signal decreases revealing a reduction in oxygen vacancies within the HZO. \fig{Fig_2}(a) plots representative spectra taken with applied laser doses that vary by 4$\times$, where the signal is found to be larger for lower doses after normalization.  The inverse relationship between the PL-signal and laser dose exhibits a near exponential trend that extends over almost two decades of laser dose [see \fig{Fig_2}(b)]. Despite the significant change in the strength of the PL-signal, the spectral profile remains nearly constant. The nearly constant spectral profile demonstrates that the energies of the trap states giving rise to the PL signal remain relatively constant.  By extension, this implies that the oxygen vacancies\textendash and their accompanying local environments\textendash are themselves not being significantly altered by the laser but that their concentration is primarily changing.  

Oxygen vacancy concentration can be spatially patterned in direct proportion to the laser dose, as evidenced by induced changes in the PL-intensity.  \fig{Fig_2}(c) shows a single PL-image in which regions subjected to doses of 0.77 mJ/$\mathrm{\mu m^{2}}$, 0.46 mJ/$\mathrm{\mu m^{2}}$, and 0.15 mJ/$\mathrm{\mu m^{2}}$ were analyzed.  Rectangular and ``L"-shaped contours are clearly evident.  These regions map directly to the overlap between successive PL-scans shown schematically in \fig{Fig_2}(d) and underscore that the laser is the primary cause of the change in PL-intensity. In addition, the relatively constant intensity within a given region qualitatively indicates that spatial variation of oxygen vacancy concentration within the film itself is relatively minor. Transition regions between the differing ranges have a width of $\sim 2$ \mum, which is on the order of 4$\times$ the beam diameter.  As no attempt was made to control the transition width, it is by no means a lower limit but rather provides further evidence that the light-matter interaction is driving the change. 

\begin{figure}[htbp]
\centering
\includegraphics[width=88 mm]{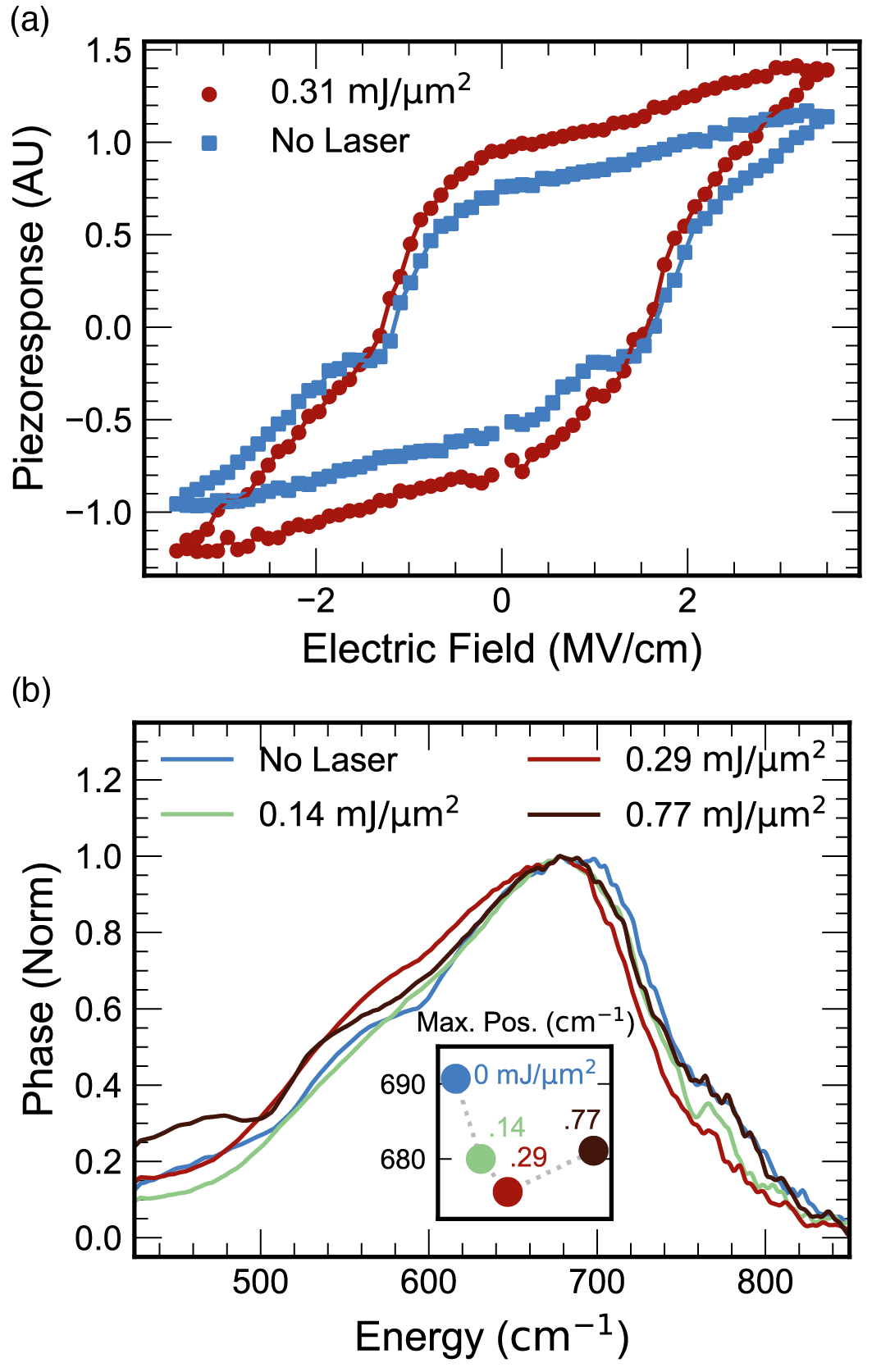}
\caption{(a) Hysteresis loops acquired using band excitation piezoresponse force microscopy in locations with and without laser exposure showing reduced pinching with exposure at the reported level of 0.31 mJ/$\mathrm{\mu m^{2}}$.  These changes are caused by promotion of the ferroelectric phase as determined by the (b) changes in the normalized phase signal from infrared nanospectroscopy (\ie nanoFTIR).  The shift of the signal to lower energies and asymmetric broadening observed at moderate laser dosing (0.14 and 0.29 mJ/$\mathrm{\mu m^{2}}$) are characteristic of an increase in the ferroelectric (\pca) phase. Inset: Location of maximum deduced by fitting the main spectral feature in the range of 600 to 750 \cm using the sum of two Lorentzian distributions.}
\label{Fig_3}
\end{figure}

As has been established, oxygen vacancies are a strong factor dictating the stabilization of the ferroelectric orthorhombic (\pca) phase in \ce{Hf_{0.5}Zr_{0.5}O_2}.\cite{jaszewski_2022,mittmann_2020}  The laser-based and spatially patterned changes in oxygen vacancy concentration therefore impact the phase composition of the \hzo films and thus their ferroelectric and electromechanical responses. To support this assertion, regions of varying laser dose were examined with local piezoresponse force microscopy (PFM) and synchrotron infrared nanospectroscopy (\ie nanoFTIR) techniques.\cite{kelley_2023,bechtel_2020} Both methods utilize atomic force microscopy (AFM) by using a nanometer size probe to spatially interrogate the local ferroelectric and structural properties, respectively. Briefly, we deploy band excitation PFM (BE-PFM) to assess the local ferroelectric properties of \ce{Hf_{0.5}Zr_{0.5}O_2}. Band-excitation PFM applies a defined band of excitation frequencies to the AFM cantilever and measures the resulting amplitude and phase of the electromechanical response (\ie piezoresponse), which can be related to the ferroelectric polarization orientation.  Similarly, in nanoFTIR, infrared light is scattered off an AFM cantilever providing spatially resolved information of the infrared response, which is dependent on the structural phase. Successive measurements with nanoFTIR in the same region exhibited no substantive difference indicating that the lower-energy infrared irradiation did not change the oxygen vacancy concentration like the visible laser.  Methods used to obtain this data are analogous to previous reports.\cite{jaszewski_2023,kelley_2023} Band excitation PFM in conjunction with nanoFTIR provides fundamental insights into the ferroelectric properties correlated with structural information. 

Results of these measurements are provided in \fig{Fig_3} where changes in piezoresponse and phase make-up are observed with laser dose. The relatively square PFM-loops exhibiting saturation in \fig{Fig_3}(a) highlight that the as-deposited (\ie ``No Laser") films contain at least some fraction of the ferroelectric \pca~ phase.  Pinching of the hysteresis response is clearly visible, however, as is common for films like these that have yet to undergo the ``wake-up" process. Upon laser irradiation, this pinching is diminished from the hysteresis loop while the piezoresponse slightly increases resulting in a more traditional ferroelectric hysteresis loop like that observed as \hzo based devices wake-up.\cite{starschich_2016,fields_2021c}  Wake-up has been attributed to a redistribution of oxygen vacancies. Here, the redistribution has been accomplished via the photoinduced reduction in their concentration. 

Control of the oxygen vacancy concentration to an optimal level promotes stabilization of the ferroelectric phase.\cite{jaszewski_2022,mittmann_2020} Thus, laser-induced changes in the oxygen vacancy concentration can be used to promote the ferroelectric phase consistent with a reduced pinching of the piezoresponse in \fig{Fig_3}(a).  The infrared spectra shown in \fig{Fig_3}(b) support this conclusion, where the normalized phase signal is used as a surrogate for absorption to assess changes in phase fraction with laser exposure.\cite{bechtel_2020,govyadinov_2013,jaszewski_2023} Moderate laser dosing (\eg 0.14 and 0.29 mJ/$\mathrm{\mu m^{2}}$) induces a shift in the primary phase signal from 700 \cm to 680 \cm accompanied by an asymmetric broadening to the low-energy side. The magnitude of this shift is highlighted within the inset to \fig{Fig_3}(b), where the fitted position of the primary feature between 600 to 750 \cm changes non-monotonically with laser dose. The initial red-shift in the main feature accompanied by the broadening are each consistent with an increasing amount of the ferroelectric orthorhombic phase (\pca) present in the film that emerges in response to the laser dosing.\cite{jaszewski_2023}  From this observation, we deduce that the reduction in vacancy concentration induced by laser exposure has, therefore, promoted the creation of the ferroelectric orthorhombic phase and\textemdash in conjunction with any redistribution of oxygen vacancies with the change in concentration\textemdash the piezoresponse of the film [see \fig{Fig_3}(a)].

The promotion of the ferroelectric \pca~ phase with laser exposure has a finite limit, however.  There is a tipping point in the ''Goldilocks problem" by which the laser-induced reduction of oxygen vacancies becomes ``too much" and promotes the emergence of the monoclinic rather than the ferroelectric phase.  The phase signal at the highest examined laser dose of 0.77 mJ/$\mathrm{\mu m^{2}}$ shows this crossover in its spectrum [see \fig{Fig_3}(b)].  Specifically, the spectrum reverts back to exhibiting features observed in the ``No Laser" condition as evidenced by a shifting to higher energy of the most intense feature [see inset to \fig{Fig_3}(b)] and a re-emergence of the mode near 780 \cm as well as those at 450 and 540 \cm.  Together, these spectral characteristics are indicative of the equilibrium monoclinic ($P2_1/c$) phase.\cite{jaszewski_2023} There is, therefore, a ``sweet spot" in laser dosing for ferroelectric phase promotion.

Photon absorption alters the local chemical potential of the HZO.  It can also  modify any oxygen containing adsorbate on the surface.\cite{stagi_2015,glass_2019}  These changes together can promote the incorporation of oxygen into the lattice. The incorporation of oxygen into the lattice\textemdash or alternatively a reduction in the oxygen vacancy concentration\textemdash can drive a phase transformation in HZO.  Here, these mechanisms have been leveraged to promote the ferroelectric phase in \hzo thin films by modest ($\sim$0.30 mJ/$\mathrm{\mu m^{2}}$) post-deposition laser dosing with 488 nm (2.54 eV) light, as confirmed using a combination of piezoresponse force microscopy and nanoFTIR measurements.  These changes can be spatially patterned and thus provide a spatially controlled, post-synthesis, low-entry method to tailor phase in \hfo-based systems.


\section*{Acknowledgements}
S.T.J. acknowledges support from the U.S. National Science Foundation’s Graduate Research Fellowship Program under grant DGE-1842490. Optical characterization, electrical characterization, and analysis were supported by the Center for 3D Ferroelectric Microelectronics (3DFeM), an Energy Frontier Research Center funded by the U.S. Department of Energy, Office of Science, Basic Energy Sciences under Award No. DE-SC0021118. AFM experiments were conducted as part of a user project at the Center for Nanophase Materials Sciences, which is a U.S. Department of Energy User Facility at Oak Ridge National Laboratory. NanoFTIR was conducted at beamline 2.4 of the Advanced Light Source. This research used resources of the Advanced Light Source, which is a DOE Office of Science User Facility under contract no. DE-AC02-05CH11231. The authors acknowledge beamline scientists Hans Bechtel and Stephanie Gilbert Corder for their technical assistance in nanoFTIR measurements. This paper describes objective technical results and analysis. Any subjective views or opinions that might be expressed in the paper do not necessarily represent the views of the U.S. Department of Energy or the United States Government.

\section*{Data Availability}
The data that support the findings of this study are available from the corresponding author upon reasonable request.

%


\end{document}